\documentclass[onecolumn,superscriptaddress,amsmath,amssymb,floatfix,aps,showpacs]{revtex4}
\usepackage{amsmath}    % need for subequations

\usepackage{graphicx}   % need for figures
\usepackage{verbatim}   % useful for program listings
\usepackage{color}      % use if color is used in text
\usepackage{subfigure}  % use for side-by-side figures
\usepackage{hyperref}   % use for hypertext links, including those to external documents and URLs
\usepackage[symbol]{footmisc}

\usepackage{amsmath}    % need for subequations
\usepackage{hyphenat}
\usepackage[normalem]{ulem}
\usepackage{array}

\usepackage{color}      % use if color is used in text
\usepackage{soul}
\usepackage[dvipsnames]{xcolor}

\begin{document}
%\bibliographystyle{prsty} % Choose Phys. Rev. style for bibliography

%\title{Machine Learning for  Gene Regulatory Network}

% proposte jonny per titolo
\title{Prediction of gene expression time series and structural analysis of gene regulatory networks using recurrent neural networks}
%\title{A parallel dual attention recurrent neural network predicts gene expression time series and discriminates gene regulatory network architectures}

%\title{Data-Driven model to infer Gene Regulatory Networks and cancer formation}
\author{Michele Monti\footnote[1]{These authors contributed equally to this work.}\footnote[2]{michele.monti@iit.it; gian.tartaglia@iit.it}
} 
\affiliation{Centre for Genomic Regulation (CRG), The Barcelona Institute for Science and Technology, Dr. Aiguader 88, 08003 Barcelona}
\affiliation{RNA System Biology Lab, department of Neuroscience and Brain Technologies, Istituto Italiano di Tecnologia, Via Morego 30, 16163 ,Genoa, Italy.}
\author{Jonathan Fiorentino$^{\star}$} 
\affiliation{Center for Life Nano- \& Neuro-Science, Istituto Italiano di Tecnologia, Viale Regina Elena 291, 00161, Rome, Italy} 
%\\ RNA System Biology Lab, department of Neuroscience and Brain Technologies, Istituto Italiano di Tecnologia, Via Morego 30, 16163 ,Genoa, Italy?}
\author{Edoardo Milanetti} 
\affiliation{Department of Physics, Sapienza University, Piazzale Aldo Moro 5, 00185, Rome, Italy}
\affiliation{Center for Life Nano- \& Neuro-Science, Istituto Italiano di Tecnologia, Viale Regina Elena 291, 00161, Rome, Italy} 
\author{Giorgio Gosti} 
\affiliation{Center for Life Nano- \& Neuro-Science, Istituto Italiano di Tecnologia, Viale Regina Elena 291, 00161, Rome, Italy} 
\author{Gian Gaetano Tartaglia$^\dagger$} 
\affiliation{RNA System Biology Lab, department of Neuroscience and Brain Technologies, Istituto Italiano di Tecnologia, Via Morego 30, 16163 ,Genoa, Italy.}
\affiliation{Center for Life Nano- \& Neuro-Science, Istituto Italiano di Tecnologia, Viale Regina Elena 291, 00161, Rome, Italy} 
\affiliation{Department of Biology and Biotechnology Charles Darwin, Sapienza University of Rome, Rome 00185, Italy}

\begin{abstract}
Methods for time series prediction and classification of gene regulatory networks (GRNs) from gene expression data have been treated separately so far. The recent emergence of attention-based recurrent neural networks (RNN) models boosted the interpretability of RNN parameters, making them appealing for the understanding of gene interactions. In this work, we generated synthetic time series gene expression data from a range of archetypal GRNs and we relied on a dual attention RNN to predict the gene temporal dynamics. We show that the prediction is extremely accurate for GRNs with different architectures. Next, we focused on the attention mechanism of the RNN and, using tools from graph theory, we found that its graph properties allow to hierarchically distinguish different architectures of the GRN. We show that the GRNs respond differently to the addition of noise in the prediction by the RNN and we relate the noise response to the analysis of the attention mechanism. In conclusion, this work provides a a way to  understand and exploit the attention mechanism of RNN and it paves the way to RNN-based methods for time series prediction and inference of GRNs from gene expression data.
\end{abstract}

\maketitle

\section{Introduction}
Recent technological innovations, such as chromatin immunoprecipitation sequencing (ChIP-seq) and RNA sequencing (RNA-seq), allow the systematic study of  complex networks formed by  interactions among proteins with DNA and RNA \cite{Vidal2011}. These  approaches are strongly advancing system biology and  open up new opportunities in medicine, allowing us to study complex diseases that affect several genes \cite{Barabasi2011,DimitrakopoulouKonstantina2014}.
\\ \\ 
The complex interaction network among genes in a cell form the gene regulatory network (GRN). In a GRN, there are typically transcription factors (TFs) proteins  and target genes. TFs activate and inhibit the transcription of target genes. In turn,  target genes produce other TFs or proteins that regulate cell metabolism \cite{Monti2021}. In this context, it is fundamental to obtain an accurate picture of the interactions among TFs and their target genes, which is known as the GRN inference problem. For this reason, we investigated how Deep Neural Networks (DNN) can be exploited to classify GRNs considering different network topologies.
\\ \\
Predicting the behaviour of a stochastic network is an extremely challenging task in modern science. Indeed, the measurable output of many processes, such as the evolution of the stock market, meteorology, transport and biological systems, consists of stochastic time traces \cite{wang2016financial, ouma2021rainfall, zhang2020temperature, raeesi2014traffic, panella2011advances}. Inferring information on the behaviour of these systems gives the possibility to predict their future changes, in turn allowing to have a thorough knowledge of the evolution of the system and, on that basis, make functional decisions. Many  mathematical methods relying on inference theory have been developed, which often are meant to learn mean properties of the systems and guess the future behaviour of the various system-specific variables \cite{Minas2017}. 
Knowing the actual future behaviour of a stochastic system allows to infer parameters of a mathematical model that is suited both to describe it and to predict the evolution of data as well as possible. These models can be  defined analytically, but the number of parameters to infer grows exponentially with the size of the network, making the system difficult or even impossible to learn \cite{Boutaba2018}.
\\ \\ 
Analytical models give the possibility of inferring the physical parameters of the system, or at least parameters that have a well-defined meaning for the system. This aspect plays a crucial role because it allows to obtain a physical interpretation of the network due to the interaction between its constituent elements. In this sense, a learning process is not only able to estimate the prediction of temporal evolution data, but it also provides the generative model which is, in some cases, physically interpretable.
Somehow parallel to the general problem of predicting the dynamical evolution of a time sequence, there is the task of classifying a set of time series. In this context, the estimated parameters obtained from the predictive model can be used in a classifier to classify different time series \cite{Langkvist2014}.
Among several predictive methods based on machine learning techniques, Neural Networks (NN) have undergone extensive development in recent years. In particular, DNN are powerful mathematical architectures that are well suited to perform tasks where ordinary machine learning procedures on a given analytical model fail or cannot be used \cite{Muzio2021}. 
However, the parameters of DNNs do have not have a straightforward physical interpretation, meaning that the model is able to predict the dynamic of the system, but it does not provide direct information about the variables.
DNN perform a set of vector-matrix and matrix-matrix linear operations, concatenated via other linear or non-linear transformations, in  order  to  give  a  certain  output  given  an input  vector.   
One of the key points in using DNNs is the process of setting parameters, which depends both on learning the network from the input data and on the selected architecture.
Among the various types of Neural Networks, Recurrent Neural Networks (RNN) are best suited for the prediction of time series data~\cite{Che2018}. Basically, these networks have a layer that is used to feed the network at each time step together with the input vector, conferring a  sort of memory to the network. 
In this sense, it is possible to shape the network architecture relying on theoretical analysis on the power of predicting time varying signal taking information
from the past.  
The amount of information from the past in a model, as well as how it is processed,  depends on the type of problem we have to solve. For instance, in order to predict the next time step of a time sequence of a noisy process we can think that we do not need information from far in the past, evermore it could be redundant and misleading.
\\ \\ 
In this work, a DNN approach was adopted with the aim of inferring information on the behaviour and on the structure of GRNs. To evaluate the effectiveness of this approach, we generated a synthetic gene regulatory model and its stochastic evolution in time, and we use a deep neural network to predict the future behaviour of the time traces of the system. 
Through a series of statistical analyses of the results obtained, we took into account the DNN parameters in order to infer information on the physical structure of the gene interactions. We show that this approach can be used to categorize protein expression profiles obtained from GRN
with different topological structures. \emph{e.g.} master regulator genes or oscillating gene networks.
\\ \\
At the heart of our approach there is the consideration that experiments, such as RNA-seq, can produce time traces of gene expression, but they do not provide any information on the interactions among genes \cite{Wang2009}. Knowing the structure of the underlying gene regulatory network is crucial to understand how the system would respond to perturbations, such as mutations or new external inputs. This is particularly relevant for the study of complex diseases, for which the genes involved are known. Among all, various types of cancer are characterized and described with this approach, since they often originate from a modification or an alteration of the gene regulatory network that governs a given cellular task \cite{Wang2009}. 
Therefore, inferring the physical structure of a GRN from the time behaviour of its output is a key problem in biology, and the ability of having a control over it could open the doors to a broader understanding of the biological mechanisms at the origin of pathologies, thus suggesting new strategies from a diagnostic and therapeutic point of view.
\\ \\ 
In this work we used the Dual Attention mechanism (AM) that has been recently introduced \cite{Lindsay2018}. Using this approach it is possible to accurately predict the next time step of a time series relying on the previous time points. The idea is that from an endogenous set of variables evolving in time, it is possible to predict the behaviour of a target variable. More formally, in a stochastic network that evolves in time, to predict the state of the $i-th$ variable, the previous T time steps of the ensemble of variables that are likely responsible for its evolution need to be considered.  
Moreover, the dual attention mechanism is needed to infer the functional interaction among the variables. The first AM selects the genes out of the pool that are master regulators
of the given state. The second AM is applied on the selected genes of the first level and it prioritizes important time steps of their expression.
These two attention mechanisms are integrated within an LSTM-based recurrent neural network (RNN) and
are jointly trained using standard back propagation.  In this way, the Dual Attention-Recurrent Neural Network
(DA-RNN) will adaptively select the most important time steps of most relevant inputs and capture the long-term temporal dependencies. Since our goal is to predict the dynamic of all the genes that belong to the GRN, we build up a parallel scheme of DA-RNN.
We use RNN because we know that that this neural networks can store several patterns and that the network structure affects the dynamics and the number of the stored patterns \cite{Folli2018,Leonetti2020,Gosti2019}.
\\ \\
To evaluate the general usefulness of the DA-RNN, we generated synthetic time series of gene expression from different classes of GRNs, resembling known biological networks. We trained a parallel DA-RNN for each GRN and we showed that it predicts the future behaviour of the time traces with high accuracy. Next, we focused on the input attention mechanism, with the goal of gathering information about the structure of the different GRNs used. Relying on graph theory and network analysis, we studied different properties of the attention layer, finding that in general they are able to discriminate different GRN architectures, thus identifying a structural information about the GRN in the attention mechanism of the DA-RNN. We observed that the robustness of the prediction by the DA-RNN to noise addition is different for the various GRN architectures, and we compared it to the properties of the input attention. Our work represents a first promising application of a DA-RNN to the prediction of time series gene expression data and it provides a step beyond in the analysis of the interpretability of the attention mechanism of deep neural networks.
 
%%%%%%%%%%%%%%%%%%%%%%%%%%%%%%%%%%%%%%%%%%
\section{Materials and Methods}

\subsection{Data generation, the Gillespie Algorithm}

The mean dynamics of a GRN can be described as a system of chemical rate equations \cite{Barbuti2020, tkacik2012}, but biological systems are intrinsically stochastic. The key connection between the chemical rate equations and the stochastic characterization  follows from Eqns. \eqref{eqProb} and \eqref{eqMean}, where we show how the rate of probability of each transition is directly connected to the parameters from a system of differential equations. 
In this paragraph, we briefly describe how to propagate the stochastic dynamic of the gene regulatory system. We simulate the stochastic dynamic in order to generate the data suitable for the inference methods that we are going to analyse.

The Gillespie algorithm is an exact stochastic simulation procedure for a set of Poissonian events connected among each other. The algorithm, introduced by D.J. Gillespie in the 1970, accurately simulates mixed systems of chemical reactions. Indeed, in a well mixed system the diffusion is neglected, considering for the computation that at each time step all the molecules can be in contact with each other.
In this context, each reaction of the system is assumed to follow the Arrhenius law, thus the probability that a reaction $j$ happens in a time $\tau$ is
\begin{equation}
P(j,\tau) = e^{-a_{j} \tau}~~,
\end{equation}
where $a_j$ is the propensity of the reaction $j$ that, in the linear case, is just $k_j x_j$, where the $k_j$ and $x_j$ are the rate and the substrate of the reaction. 
Considering a set of independent processes, the probability of having a certain reaction after time $\tau$ is given by
\begin{equation}
P(\tau) = \prod_j e^{-a_{j} \tau} = e^{-\sum_j a_j \tau} = e^{- a_{tot} \tau}~~,
\end{equation}
where $a_{tot} = \sum_j a_j$ is the total propensity to react of the system. This defines an exponential probability distribution of having one reaction after time $\tau$.
The next step is to define which reaction will happen; this is simply obtained from the relative propensity of each reaction
\begin{equation}
P(j)= \frac{a_j}{a_{tot}}~~.
\end{equation}

The steps of the algorithm are the following:

\begin{enumerate}
\item Define the network, the matrix of interaction, the nature of the interactions, the rates of the reactions, the number of species (nodes), the final time $T_{max}$ of the simulation and the propensity functions $a_j$;
\item Select the time of the first reaction from the exponential probability distribution of the times: $P(\tau) = e^{- a_{tot} \tau}$;
\item Select the reaction from the relative propensity: $P(j)= \frac{a_j}{a_{tot}}$;
\item Upgrade the species according to the rules of the reaction $j$;
\item Compute the new propensities $a_j$ and $a_{tot} = \sum_j a_j$ from the updated set of the network;
\item Update the time $t +=\tau$
\item Repeat from step 1 until $t<T_{max}$~~.
\end{enumerate} 

For a system of interacting nodes where the statistical structure does not take into account diffusion but just the reaction rates, this algorithm gives exactly the real time traces of the status of the nodes involved. We stress that the time step of the simulation is not constant, but being a simulation of a Markov process, where the state of the network at time $T+\tau$ just depends on the state at time $t$, we can sample the state of the network for any $dt$.

%Using this algorithm we want to simulate the dynamic of a GRN, we will use a Boolean approach to optimise the performance of the RNN that needs to predict the time behaviour of the system.	\gio{[Boolean?]}

\subsection{Dual attention mechanism structure and learning procedures}
\label{train}

Instead of using a classical neural network approach, we propose a new method to predict the concentrations of a set of proteins measured at times $<t$ . The method uses state of the art Deep Learning methodologies including Long Short-Term Memory (LSTM) and attention mechanism (AM).

We have implemented a Deep Neural Network model that relies on the Dual-Stage Attention-Based Recurrent Neural Network (DA-RNN) for Time Series Prediction \cite{qin2017dualstage}, which has been developed in order to predict the behaviour of a target quantity in a stochastic ensemble of time-traces. 
The network has an Encoder-Decoder structure and, in both parts, an attention layer is the central core. The encoder attention layer weights the $N$ time traces to optimise the prediction of the target, while the decoder attention layer looks for the best way of weighting the past time points in each time trace. The combination of these two layers finds two key characteristics of the prediction for a given target time trace: on one hand, it finds the most important elements (in our case, genes) in the system that should be considered for predicting the target and, on the other hand, it quantifies the importance of the different past points. 

We are interested in predicting the further time step of the whole network of interacting genes. To this end, we implemented the parallel scheme shown in figure \ref{fig2}. Given the input data, $N$ parallel DA-RNN are trained, where the $i-th$ network predicts the behaviour of the $i-th$ target gene. We hope to find encoded in the attention layer functional information about the interaction among the genes.

To optimize  network performance, a standard scaling of the data is done, converting all the time traces to zero mean and unit variance.

We use minibatch stochastic gradient descent (SGD) together with the Adam optimizer \cite{kingma2017adam} to train the model. The size of the minibatch is 128. The cost function of the back propagation is the mean square error (MSE):
\begin{equation}\label{mse}
    MSE(y_p, y_c)= \sum_{i=1}^{M} \left(y^{(i)}_p(t)-y^{(i)}_c(t)\right)^2~~,
\end{equation}
where $y_p$ and $y_c$ are the vectors of the prediction and the target data, respectively, and $M$ is their length.

We have implemented the network in python using the py-torch library.

\subsection{Deep Neural Network Parameter Analysis}

To extract the weights of the input attention mechanism for each GRN, we trained a DA-RNN for each target gene, as described in the previous section, and we collected the vector of $N$ input attention weights, where $N$ is the number of genes. This results in the definition of a matrix $\mathbf{A}=\{a_{ij}\}$ for each GRN, where the element $a_{ij}$ is the input attention weight of gene $j$ obtained by training the neural network on target gene $i$. Next, we treated the input attention matrix as a graph and we computed several network properties. In order to investigate the organization of the interactions that make up the network architecture, we choose three different local parameters, which are defined for each node of the network. The first is the clustering coefficient, which describes the topological/structural organization of the interactions involving a node and its first neighbors. As a second descriptor we consider the betweenness centrality, which is a descriptor of the role of each node in relation to all the minimum cost paths that pass through the given node and that connect every other pair of nodes in the graph. The last descriptor, the hubscore, is directly related to the intrinsic properties of the adjacency matrix, since it is defined as the principal eigenvector of the adjacency matrix of the graph.
To this end we use the "transitivity" function, "betweenness" function and "hubscore" function of the "igraph" package of R (The Igraph Software Package for Complex Network Research. InterJournal 2006, Complex Systems) for the three descriptors respectively. For each case we consider a weighted graph. 
More specifically, the three network descriptors are defined as follows:\\
(i) The clustering coefficient is given by:

%weighted C_i = 1/s_i 1/(k_i-1) sum( (w_ij+w_ih)/2 a_ij a_ih a_jh, j, h)

\begin{equation}
    C_i = \frac{1}{s_i} \frac{1}{(k_i - 1)} \sum_{j,h} \frac{(w_{ij} + w_{ih})(a_{ij} a_{ih} a_{jh})}{2}~~,
\end{equation}

where $s_{i}$ is the strength parameter of vertex $i$, which is defined as the sum of all weighed links. The $a_{nm}$ are elements of the adjacency matrix (with both $n$ and $m$ are $i$,$j$ and $h$ indices). Finally, $k_{i}$ is the vertex degree (the number of links), $w_{nm}$ are the weights.

(ii) the betweenness centrality is defined as follows:

\begin{equation}
    b_i = \sum_{k \not = i,l \not = i} \frac{s_{kl}(i)}{s_{kl}}~~,
\end{equation}
where $s_{kl}(i)$  is number of  shortest paths ($s$) that go from node $k$ to node $l$ passing through node $i$ and $s_{kl}$ is the total number of shortest paths from node $k$ to node $l$.

(iii) The Kleinberg's hub centrality score, or hub scores, of the vertices are defined as the principal eigenvector of $AA^t$, where A is the adjacency matrix of the graph.

\subsection{Response of the prediction to noise addition}

In this section we describe the analysis of the response to the prediction by the DA-RNN to noise addition on the protein concentration. For each target gene of a GRN, we used the parameters of the DA-RNN network, trained as discussed in Section \ref{train}, to predict its level at time $t+1$ using the previous time step. With this procedure, we predict the whole time series for a given gene and we compute the mean squared error, defined in Eq. \eqref{mse}.

Then, we repeat the prediction but adding a Gaussian noise, with zero mean and variance $\sigma^2$, to the previous time step of the target gene, and we compute the MSE between the noisy prediction and the original time series. In the end, for each gene regulatory network, we obtain a matrix containing the MSE for each target gene and different values of the variance of the Gaussian. 
To study systematically if there is a different response to noise addition in the prediction of gene expression dynamic for different gene regulatory network architectures, we collected the vectors of the mean and variance of the MSE, computed over the different values of $\sigma^2$, in a matrix. For each network, the vector of the mean was ranked in increasing order and normalized to its maximum, while the variances were ranked according to the mean ranking, and they were normalized to the maximum as well. This was done because there is not a biological correspondence between the genes in the different gene regulatory networks, and to retain the relationship between mean and variance of the MSE for the same gene in each network.

\subsection{Clustering and Principal Component Analysis}

Given a matrix representing a specific network property of the input attention mechanism or the response of the prediction to noise for all the GRNs used, we performed a clustering analysis and a Principal Component Analysis (PCA) using the Python package scikits-learn.

Firstly, we scaled the data to zero mean and unit variance using the function "StandardScaler". Next, we performed agglomerative hierarchical clustering on each matrix, with complete linkage and Euclidean distance, using the function "AgglomerativeClustering". For each matrix, we chose the optimal number of clusters as that maximizing the average silhouette score. The silhouette score for a given sample is defined as
\begin{equation}
    s=\frac{D_N-D_I}{\max{(D_I,D_N)}}~~,
\end{equation}
where $D_I$ is the mean intra-cluster distance for the sample under consideration and $D_N$ is the distance between the sample and the nearest cluster that the sample is not a part of.

We also ran a PCA on the same matrix, using the "pca" function.

In order to compare the dendrograms obtained from the hierarchical clustering of the different matrices, we first converted them to Newick trees, which is a standard format for phylogenetic trees, then we used the R package "ape" to read the dendrograms and the R package "TreeDist" to compute the distance between each pair of dendrograms. In particular, we used the Information-based generalized Robinson–Foulds distance \cite{smith2020information}. The Robinson-Foulds distance counts the number of splits that occur in both trees. Its generalization measures the amount of information that the splits of two trees hold in common under an optimal matching. The optimal matching between splits is found by considering all the possible ways to pair splits between trees.

Finally, the comparison between the partitions obtained from the hierarchical clusterings was performed using the Variation of Information \cite{meilua2007comparing}, which measures the information that is gained or lost from a clustering to another of the same dataset. Given two partitions $\mathbf{C}$ and $\mathbf{C^{\prime}}$ of the same dataset, the Variation of information between them is defined as
\begin{equation}
    VI(\mathbf{C},\mathbf{C^{\prime}})= H(\mathbf{C}) + H(\mathbf{C^{\prime}}) - 2 I(\mathbf{C},\mathbf{C^{\prime}})~~,
\end{equation}
where $H(\cdot)$ is the entropy of a partition and $I(\mathbf{C},\mathbf{C^{\prime}})$ is the mutual information between the partitions.

\begin{figure}[t!]
\centering
\includegraphics[width=.8\columnwidth]{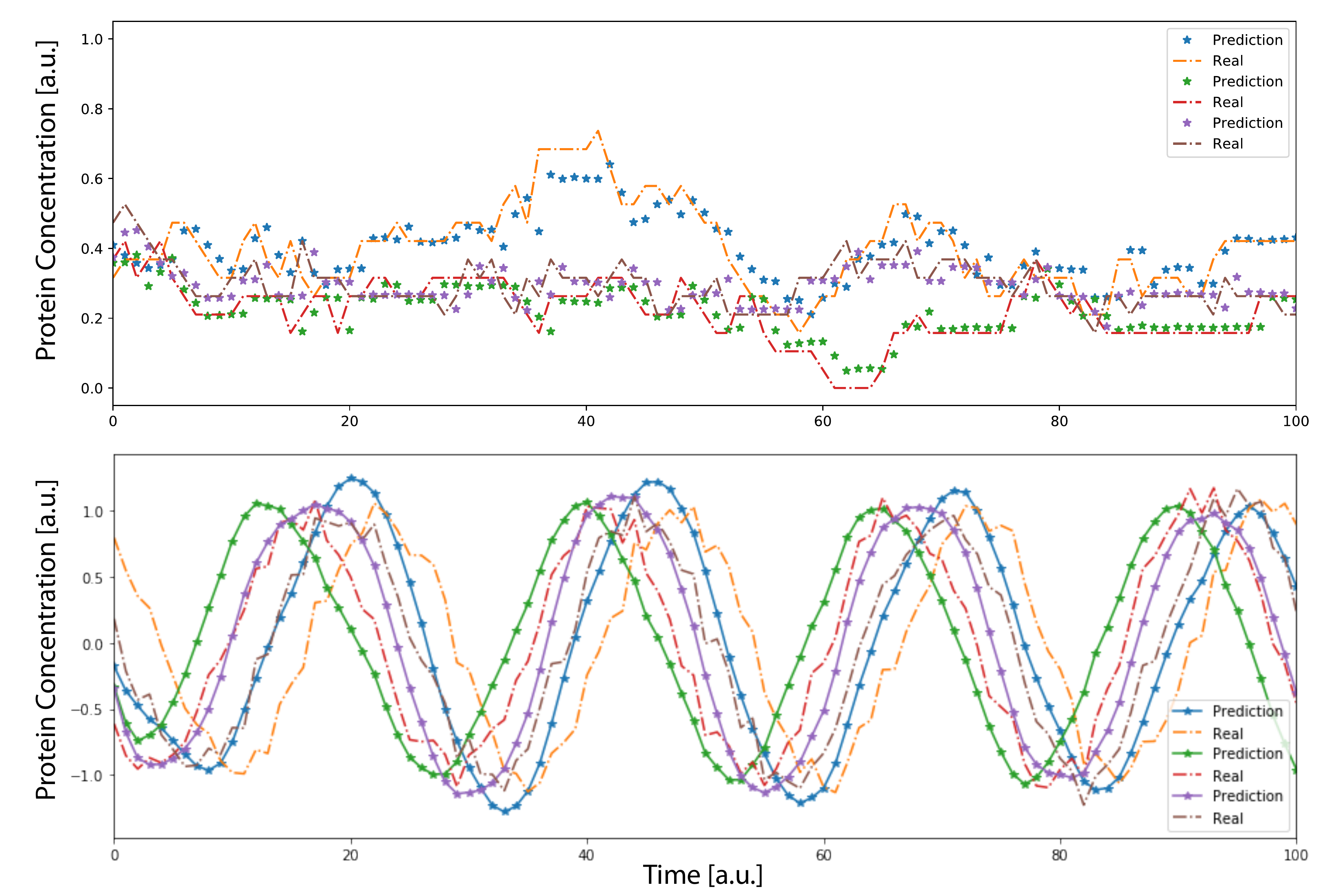}
\caption{ Gene regulatory Network dynamic with  $N=3$ genes. The dynamic is generated using a random interaction matrix $J_{ij}$. The top panel shows the continuous time trace of proteins concentration for a random GRN with $n_r=n_a = 0.2$. The bottom panel shows the oscillatory dynamic of 3 genes relative to the core clock of an oscillatory network. In both panels, dots show the prediction of the neural network implemented using the first $T =10$ time points and then letting the system evolve by its own. The stochastic data were generated using the Gillespie algorithm.
 }\label{fig:fig1}
\end{figure}

%%%%%%%%%%%%%%%%%%%%%%%%%%%%%%%%%%%%%%%%%%
\section{Results}

\subsection{Gene Regulatory Network Dynamic}

Modelling and simulating a whole gene regulatory network is a fundamental challenge for computational biophysics. Indeed, the network of interacting genes includes many chemical reactions, with the respective rates and parameters that define a high-dimensional parameter space. Moreover, the activation of a gene in the classic framework happens via a transcription factor (TF) that, in order to bind the DNA, usually goes through a set of crucial chemical post transcriptional modifications, which can drastically change the nature of the protein. 
Other physical dynamics, like tuned delay, repression, the presence of multiple TF and the cooperativity among them further complicate the endeavour of modelling gene regulatory networks.

The simplest way of describing a gene regulatory network is to define the nature of the interactions, writing the probability $P(x_i^+)$ of activating a gene $i$ expressing a protein with a concentration $x_i$ between time $t$ and $t+dt$ as:
\begin{eqnarray}\label{prob}
P(x_i^+)dt = \prod_j \left[ J_{ij}\frac{(J_{ij}+1)}{2} k_{ij} \frac{x_j^{h_{ij}}(t+\tau_{ij})}{x_j^{h_{ij}}(t+	\tau_{ij}) + K_{ij}^{h_{ij}}} + \right. \nonumber \\ \left.+ J_{ij}\frac{(J_{ij}-1)}{2} \tilde{k}_{ij} \frac{K_{ij}^{h_{ij}}}{x_j^{h_{ij}}(t+\tau_{ij}) + K_{ij}^{h_{ij}}} \right]
& \text{if} \;\;\;J_{ij} \neq 0~~,
\end{eqnarray}

where we introduced:
\begin{itemize}
\item $J_{ij}$ the interaction matrix among the genes $i$ and $j$, which is 1 if gene $j$ expresses a protein at concentration $x_j$ that will be a TF for the gene $i$, -1 if it will be a Repressor and 0 if the two genes do not interact.
\item $k_{ij}$ and $\tilde{k}_{ij}$ are the expression and repression rates for the gene $i$ from the TF (repressor) $j$
\item $h_{ij}$ the cooperativity index of the $ij$ interaction.
\item $K_{ij}$ the dissociation constant of the reaction $ij$
\item $\tau_{ij}$ the delay between the expression of the protein $x_j$ and the time when it is ready to activate (repress) the gene $i$.
\end{itemize}
Notice that the term $J_{ij}\frac{(J_{ij}+1)}{2}$ is one if $J_{ij} = 1$, and zero otherwise. While the term $J_{ij}\frac{(J_{ij}-1)}{2} $ is one if $J_{ij} = 0$, and zero otherwise.

We stress that the behaviour of the gene regulatory network described in Eq. \eqref{prob} is intrinsically dependent on the temporal dynamic of the respective number of proteins $n_{i}(t)$ expressed. For this reason, the correct way to look at that is to model the protein dynamics that, with high fidelity, mirrors the gene regulatory dynamics. Indeed, we can consider that the concentration of a protein still behaves as a stochastic process, and the probability of increase is directly proportional to the probability of expressing the correspondent gene. For our purpose, we considered that a protein can be created through the expression of a gene following the stochastic rules defined in Eq. \eqref{prob}. Moreover, we included a dilution effect on the protein concentration that leads to a constant decay rate. This definition of the stochastic process that describes the behaviour of creation and dilution of a protein completely defines our Gillespie simulations. Other effects can be taken into account, as we mention in Appendix \ref{appmeanfield}, in which we provide the computation of the mean field equations of the system.

\subsection{Recurrent Neural Network for Stochastic Time traces generated by different Gene Regulatory Network architectures}
\label{arch}

Different classes of gene regulatory networks model distinct biological systems. In this section, we describe the gene regulatory network architectures that we chose for the generation of time series data and training the deep neural network model. Firstly, we focused the analysis on the degree of connectivity, which is an important network feature. Indeed, some networks are extremely connected, meaning that the matrix $J_{ij}$ is dense, while others are sparser, i.e. the matrix $J_{ij}$ has more null values.
Here we introduce the parameters $n_r$ and $n_a$, which are the probabilities that an element of the $J_{ij}$ is -1 (inhibitory connection) or 1 (activatory connection), respectively. In other words  $n_r$ and $n_a$ multiplied by $N$ (total number of genes) can be interpreted as the mean number of repressors and activators for a given gene. Using these parameters we can tune the connectivity matrix density of the net. Indeed, if we analyze the extreme values of these parameters, for $n_r, n_a \rightarrow 1$ the net is fully connected, while for $n_r, n_a \rightarrow 0$ the net is fully sparse.
\\ \\ 
Oscillations are another important property of GRNs. Therefore, we included in our analysis oscillatory networks in which there is a central core of genes that evolve in an oscillatory fashion. The simplest way to induce oscillations in a gene regulatory system is to introduce a feed forward loop with some delay. To this end, we relied on the Goldbetter model \cite{Goldbeter1990}, where a first gene activates the expression of a second gene, which in turn activates a third one that finally represses the first node. This procedure introduces a delay in the repression of the first node, that  starts the oscillation. It can be intuitively understood thinking that gene 1 starts to rise for a certain time $\tau$ while genes 2 and 3 are getting expressed, after this time the concentration of the protein expressed by the third gene is high enough to induce the repression of gene 1, that in turn stops its expression. Consequently, the concentration of the protein goes down and the oscillation begins.
\\ \\ 
Another class of networks that we studied are those where all genes share incoming connections from a master regulator gene. In other words, we define a GRN in which the first gene $i$ shares out edges with all other genes, and in addition to these edges we add random connections among the other gene pairs.
\\ \\ 
The last class of GRNs that we studied are those connected to an external oscillatory signal. Also in this case the connectivity among the internal genes is random and quantified by the parameters $n_r $ and $n_a$ defined above.
We study how our prediction model performs and learns varying these parameters and network topology.

To predict the temporal evolution of a GRN, the neural network is trained on a certain time window of the input data, in order to predict the trend of the data for subsequent times. This time window plays a fundamental role in the training procedure. Indeed, taking it too large would provide confusing information to the network, while choosing it too short would risk losing essential information for the prediction. Clearly, using a brute force approach to find the optimal time window would be computationally expensive, so we take into account some theoretical aspects. 

In our approach we considered that the mutual information between the state of the system at time $t-\Delta t$ and the state of the system at time $t $ goes to 0 for $\Delta t \rightarrow \infty$ \cite{tostevin2009}, also for oscillatory systems \cite{monti2018} \cite{montiPRC}.
\begin{equation}
I \left[ x(t-\Delta t); x(t) \right] \rightarrow 0 \;\;\;\;\; \text{if}  \;\;\; \Delta t \rightarrow \infty~~.
\end{equation}
Moreover, previous theoretical works  quantified that the optimal way to weight the past is using the temporal autocorrelation function $C(t,\tau) = \langle x(t) x(t-\tau) \rangle_{\tau}$ \cite{tostevin2009}. This would lead us to set the time window a bit larger than the maximum autocorrelation time of the system. This information will help us in estimating this parameter and training the neural network.

We used the Parallel DA recurrent neural network described in detail in Section \ref{train}. After training, it is able to reconstruct quite well the dynamic of the system, as we can see in Figure \ref{fig:fig1} for both oscillatory and non oscillatory dynamic. 

The prediction is propagated providing to the network the first $T$ input data from the time trace and letting the DNN to predict the next time step. When the first time point (after $T$) is predicted, the network will have as the input for the next prediction the $T-1$ input data plus the first prediction point. Iterating this procedure, after a time $T$ the system would rely its prediction only on previously predicted data, showing a complete autonomous behaviour. 

\begin{figure}[t]
\centering
\includegraphics[width=.9\columnwidth]{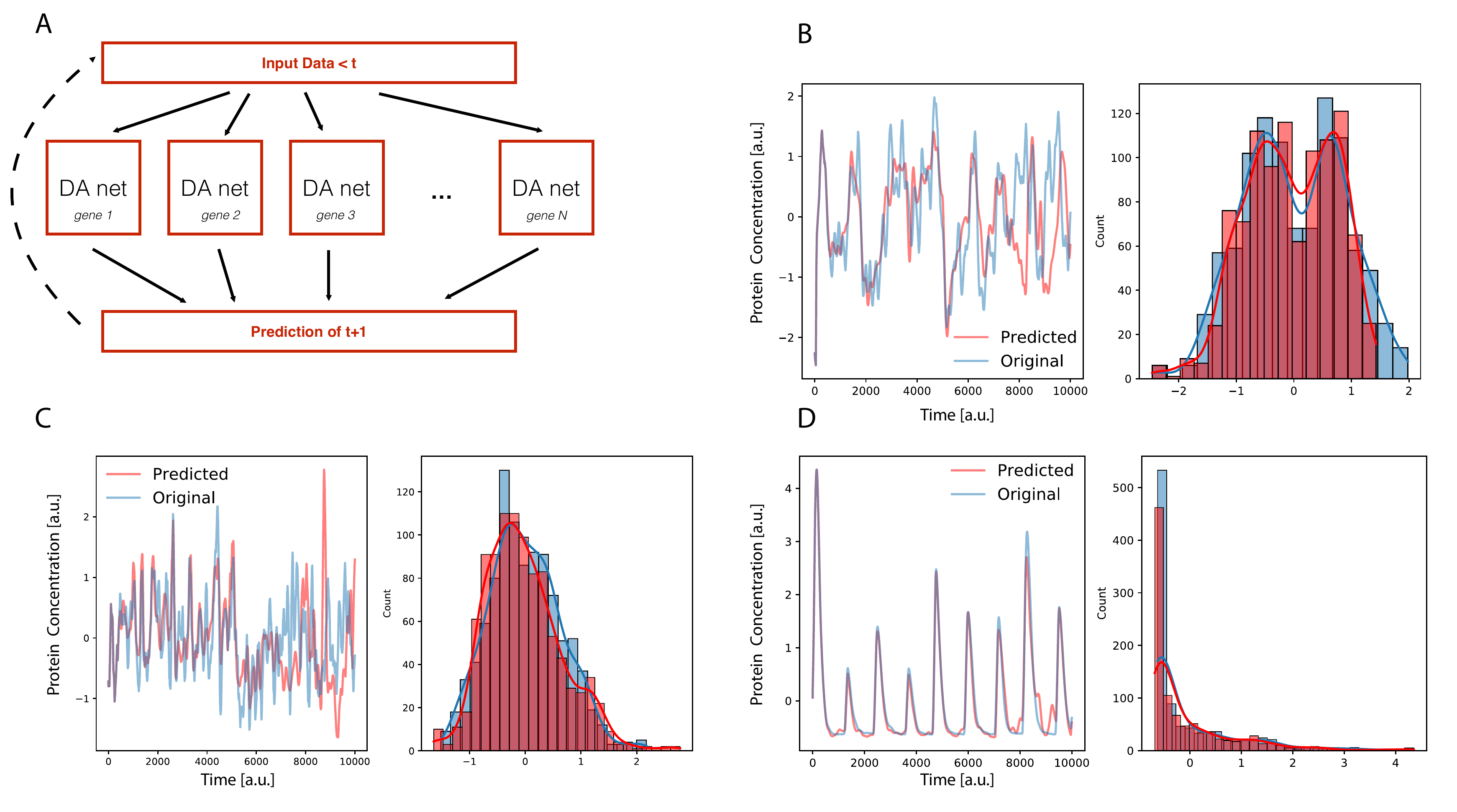}
\caption{A: Architecture of the deep neural network used in this work. Each box represents a DA-RNN devoted to the prediction of gene $i$; after the prediction of the next time step, the data are collected to fuel the inputs for the parallel nets and to predict the further time point. Panels B, C and D represent the time dynamic and the full probability function (integrated over all the time of the simulation) of having a protein at a certain concentration. We show selected example genes from an oscillatory GRN (panels B and D) and a GRN controlled by a master regulator gene (panel C), both with $n_r = n_a = 0.5$. The plots show the performance of the prediction comparing the time traces between the stochastic simulation and the DNN propagation. Moreover, the resulting probability distribution of the given protein concentration for all times is reported. We highlight that the protein concentration dynamic is reconstructed with high accuracy, and even the noise amplitude and the mean are quite well reported by the neural net. Moreover, for the genes belonging to the oscillating GRN (panels B and D) it is possible to reconstruct the amplitude and the mean of the oscillations. We stress that the dynamic shown in panel B cannot be immediately addressed to a gene driven by an oscillatory clock, but the bi-modal distribution shows that this is the case. Interestingly, this reveals that given a model trained on real data, we can look at its internal parameters that allow us to determine if the genes in the GRN belong or not to a certain network topology such as an oscillatory driven network, such as for instance the cell cycle or circadian clock.}\label{fig2}
\end{figure}

In figure \ref{fig2}B,C and D we show, for some example genes, how well the time traces and the probability distributions of protein concentration are reconstructed for different GRNs.
It is important to notice that oscillatory dynamics, which have a well characterized mean behaviour, are easily learned by the neural network, showing a very good match between the stochastic simulated data and those generated by the neural network.

\subsection{Studying the input attention mechanism and the response of the prediction to noise}
\label{att_and_noise}

The input attention mechanism of the DA-RNN assigns weights to the genes of the gene regulatory network under study, with the goal of prioritizing those that enable the prediction of the dynamic of the target gene. Thus, we reasoned that the set of input attention weights for a given gene regulatory network could reflect to some extent the structure of the regulatory interactions, allowing to distinguish different architectures of the gene regulatory networks. 
For each gene regulatory network, we extracted the input attention matrix $\mathbf{A}=\{a_{ij}\}$, as described in detail in the Methods section.

A recent study showed that the input attention mechanism of the DA-RNN does not reflect the causal interactions between the variables composing the system \cite{baric2021benchmarking}. Thus, to reveal whether there is a different structure of the input attention matrices for the GRN architectures described in Section \ref{arch}, we employed methods from graph theory.

In order to analyze the role of each network parameter for each type of network used, we need to consider the values of the descriptors for each node (gene) of the graph in order, in this case decreasing. The reason for this choice is due to the fact that there is no correspondence of a physical and / or biological nature between the genes of two different matrices, therefore there is no {\it a priori} numbering of the genes. We only evaluate two matrices that are similar to each other if the two profiles of the analyzed network property are similar. This procedure allows us to compactly describe each attention matrix with a single vector composed of a specific network property, node by node, in descending order. In particular, three local network properties have been selected for this comparison: the clustering coefficient (to have a description of the network topology given by interaction with first neighbors), the betweenness (or closeness) centrality (to describe a more complex information linked to all the possible shortest paths between the nodes) and the Hubscore (an index on the intrinsic properties of the adjacency matrix). For a deeper description of the three descriptors see the Methods section.

Next, we studied how the time series prediction by  DA-RNN is affected by noise added to the gene expression, to see if  different  regulatory architectures react in distinct ways. The addition of noise to the  time series is meant to reflect  alteration in genes belonging to  specific networks. Indeed mutations in the cell cycle genes often cause cancer by accelerating  division rates or inhibiting normal controls of the cell such as in cycle arrest or programmed  death \cite{Otto2017}. Thus, analysis of RNA-seq experiments can be used to reveal difference in the expression of cell cycle genes \cite{Caglar2020}. Other endogenous contributions to noise can be given by variability in gene expression and regulation \cite{Jong2019,Miotto2019,crisanti2018statistics}, which is already taken in account by our model, but also by noise given by the partitioning process occurring at cell duplication \cite{Peruzzi2021}. Also in non-pathological conditions, modelling the effects of noise is relevant to understand variations in experimental techniques such as RNA-seq \cite{brennecke2013accounting} and mass spectrometry \cite{du2008noise}. 

\begin{figure}[!h]
    \centering
    \includegraphics[width=\columnwidth]{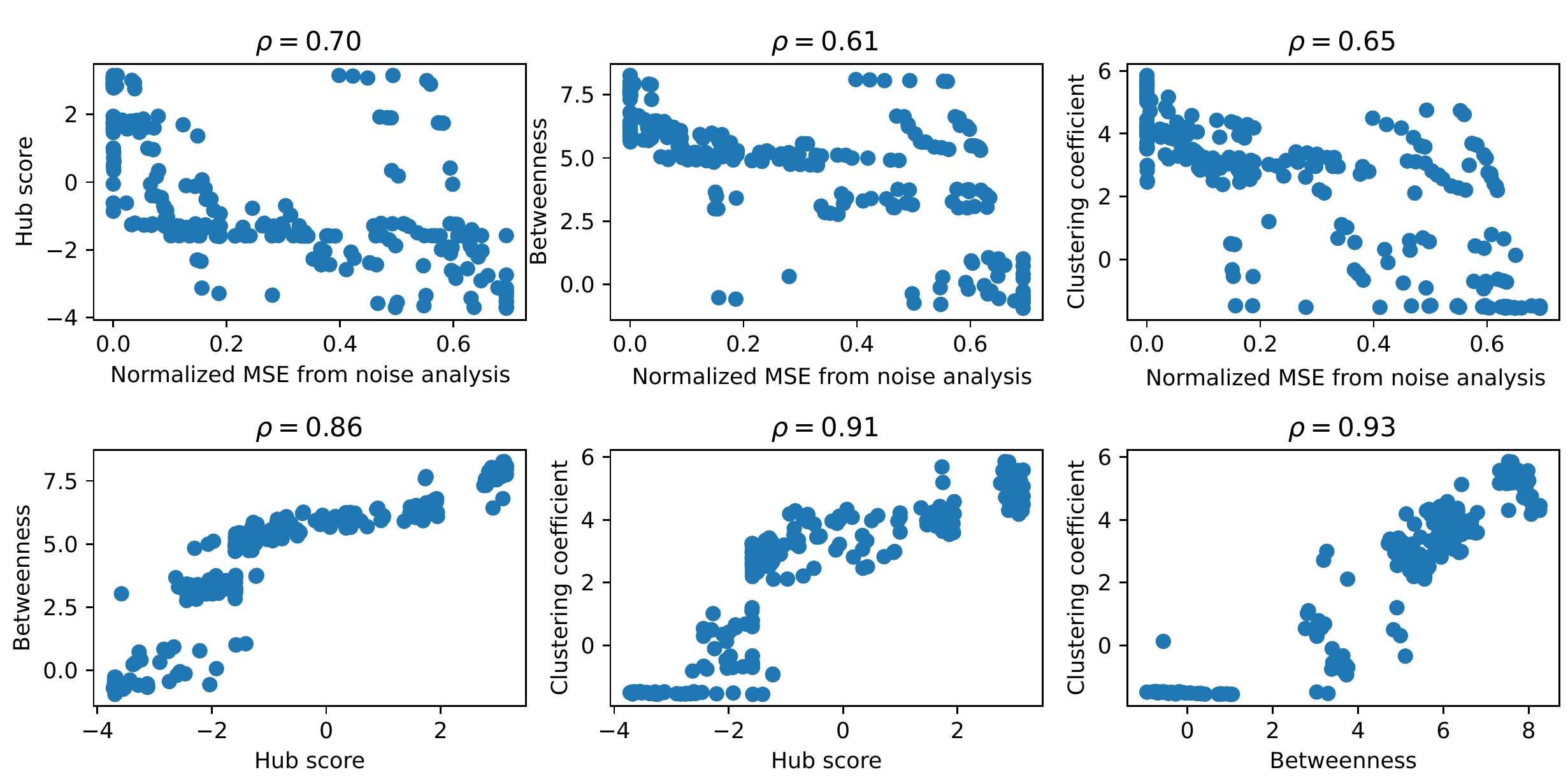}
    \caption{{\bf Comparing the matrices resulting from the input attention and response to noise analyses}. Scatter plots showing the relationship between the matrix elements of those obtained computing the network properties of the input attention (Clustering coefficient, Hub score and betweenness) and the matrix of normalized mean MSE obtained from the analysis of response of the prediction to noise. Note that the data are represented in log-log scale by using the transformation $sign(x) \ln(1+\lvert x \rvert)$, since they included both positive and negative values. The Pearson's correlation coefficient $\rho$ between each pair of properties, computed on the transformed data, is reported on top of each panel.
    }
    \label{fig:cmp_matrix}
    
\end{figure}

To this end, we used the parameters of the DA-RNN, trained as described in the Methods section, to predict the level of the target gene, but adding a Gaussian noise with zero mean and variance $\sigma^2$ at the previous time step. We computed the MSE on the prediction and we repeat the procedure for several values of $\sigma^2$. In the end, we built a matrix containing the mean and the variance of the MSE, computed over the different values of $\sigma^2$, for each gene and each GRN. We refer the Reader to the Methods section for further details. 
\\ \\ 
In order to compare the matrices obtained, we computed the Pearson's correlation coefficient between each pair of matrices; to obtain a comparable shape, for the matrix derived from the noise analysis we considered only the 
vectors of the mean MSE. The scatter plots of the matrix elements for the different properties are shown in Fig \ref{fig:cmp_matrix}: we observe that network properties reach the highest values of the correlation coefficient, while the correlation with the results from the noise analysis is lower, especially for the betweenness. 

\subsection{Network properties of the input attention distinguish gene regulatory network architectures}

\begin{figure}[!h]
    \centering
    \includegraphics[width=\columnwidth]{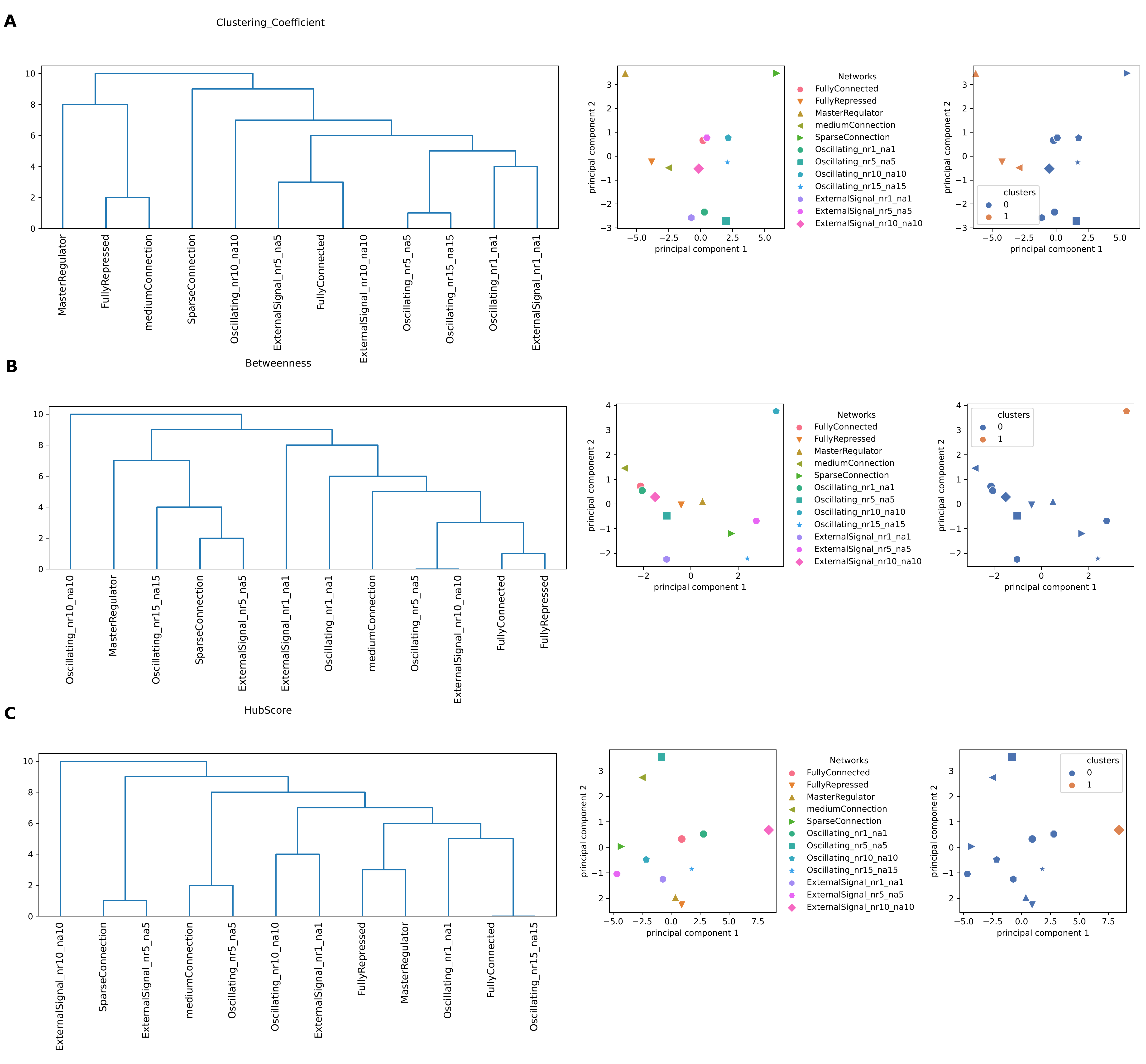}
    \caption{{\bf Clustering based on the network properties of the input attention matrices}. We show the dendrograms obtained from a hierarchical clustering (left), a PCA plot showing the 12 gene regulatory networks used (centre) and the obtained partition in groups (right) for the clustering coefficient (A), betweenness (B) and hub score (C).}
    \label{fig:clus_net_prop}
\end{figure}

To better capture  information from the attention matrices, we interpret these as a graph and use local parameters from graph theory to quantify the role of each node (gene) in the complexity of the interaction network with other nodes (genes).
To this end, we considered three local descriptors that are able to determine both local topological properties and network properties linked to the definition of the shortest path (see Methods). Using the matrices obtained for the Clustering coefficient, the Hub score and the Betweenness Centrality from the input attention matrix of each GRN, we performed a hierarchical clustering, as detailed in the Methods section. From the dendrograms shown in Fig \ref{fig:clus_net_prop} we observe a tendency in separating oscillating networks and networks controlled by an oscillating external signal from the others. This is more accentuated for the clustering coefficient, meaning that the local structural network properties, captured by the clustering coefficient as a measure of local interactions with first neighbors, keep more the information necessary to recognize GRN with different architectures. 

We ran a PCA on the same matrices and we represented the GRNs and the partition in clusters using the first two principal components (PCs), which together explain on average the $72\%$ of the variance, for each network property studied. For the clustering coefficient (Fig \ref{fig:clus_net_prop}A), the "MasterRegulator" and the "SparseConnection" networks are widely separated from the others along PC1. Regarding the betweenness (Fig \ref{fig:clus_net_prop}B), the Oscillating network with $n_r=n_a=10$ stands out along PC2, while the other networks are arranged along a continuous trajectory. Finally, for the hub score (Fig \ref{fig:clus_net_prop}C), the separation along PC1 is driven by the network with the external signal with $n_r=n_a=10$, as also reflected by the clustering shown on the right.

\subsection{Differential response to noise in the prediction of time series gene expression data}

\begin{figure}[!h]
    \centering
    \includegraphics[width=.9\columnwidth]{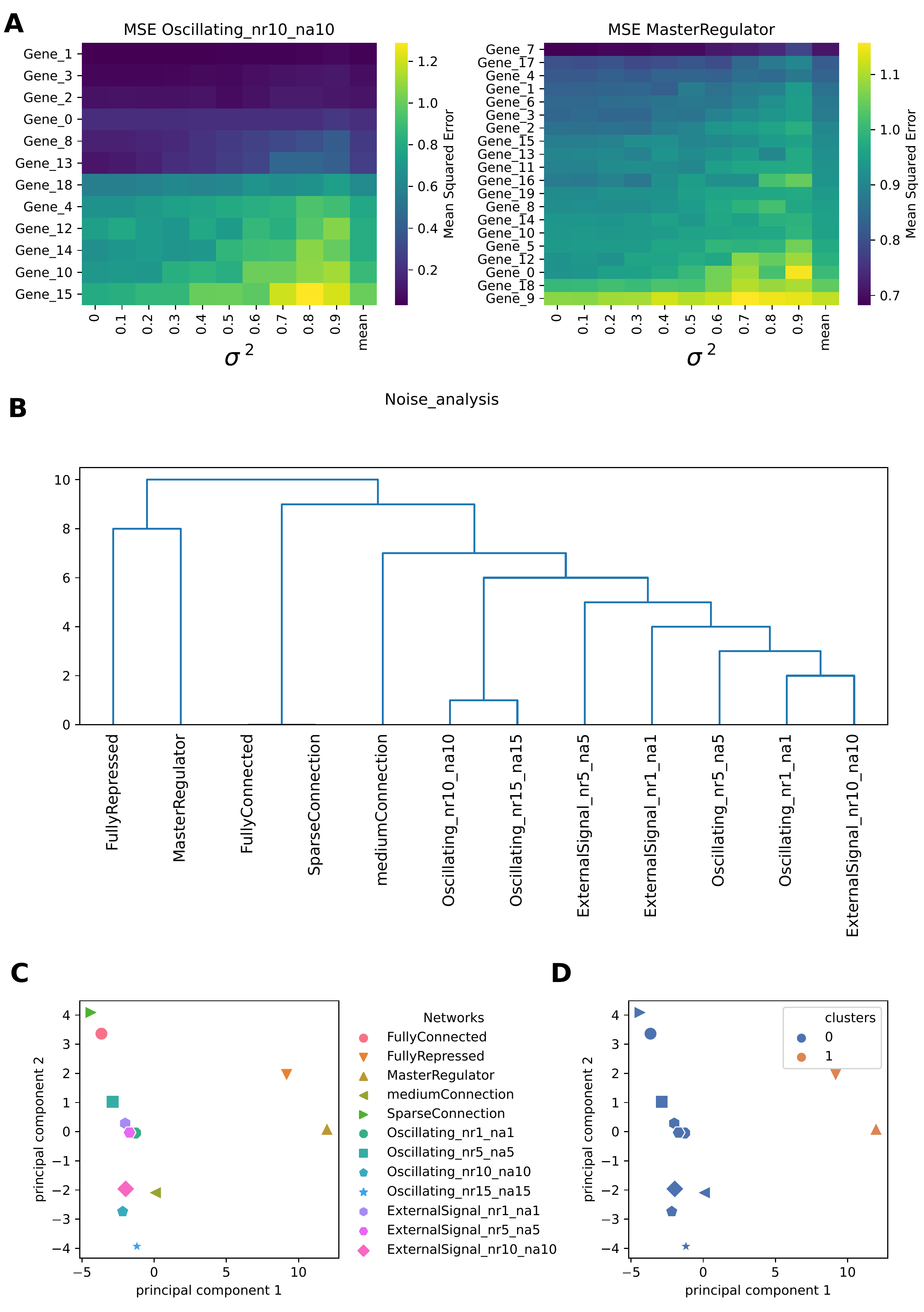}
    \caption{{\bf Impact of noise addition on time series gene expression prediction for different GRN architectures.} A: Matrices of the Mean Squared Error (MSE) on the prediction by the DA-RNN, for an oscillating GRN with 10 repressors and activators per gene on average (left) and a GRN controlled by a master regulator (right). Rows (genes) are ranked according to the mean of the MSE (last column). B: Dendrogram obtained from the hierarchical clustering of the matrix summarizing the response of the prediction to noise for each gene regulatory network. C-D: PCA showing the 12 GRNs used and their partition in clusters.}
    \label{fig:noise}
\end{figure}

In this section, we studied the matrices of the MSE obtained from the analysis of noise described in the Methods and in Section \ref{att_and_noise}. An example is shown in Fig \ref{fig:noise}A for two gene regulatory networks. Rows contain genes, ranked according to the mean over the different noise levels of the MSE (shown in the last column). As expected, we observed an increase of the MSE of the prediction increasing the variance $\sigma^2$. However, the structure of the matrices is different for different gene regulatory network architectures: for an oscillating network controlled by 4 clock genes (genes 0,1,2 and 3) the prediction by the DA-RNN is very accurate and stable against the addition of noise for some genes, while for others the MSE is larger and it increases rapidly with $\sigma^2$. Notably, the top 4 genes are the clock genes mentioned above, although the same ranking is not observed for all the oscillating networks used in this study. For the network controlled by a master regulator, the structure of the MSE matrix is more uniform, showing less differences between the genes in responding to noise addition.
\\ \\
Next, we built a matrix containing the means and the variances of the MSE for each gene and each GRN and we performed hierarchical clustering and PCA on it, as described in detail in the Methods section. The dendrogram representing the hierarchical clustering is shown in Fig \ref{fig:noise}B. We observed that oscillating networks and GRNs controlled by an oscillating external signal are clearly separated from the others. This is somewhat similar to what we observed from the network properties of the input attention, but even sharper. The "FullyRepressed" and "MasterRegulator" GRNs clearly stand out, since in these networks most of the genes randomly oscillate around zero. From Fig \ref{fig:noise}C-D we notice that PC1 reflects the top branching of the dendrogram, with the "FullyRepressed" and "MasterRegulator" GRNs widely separated from the other networks, as found also in the clustering shown on the right, while PC2 shows a finer separation between the other GRN architectures.

\subsection{Comparison between the clustering of gene regulatory networks}

\begin{figure}[!h]
    \centering
    \includegraphics[width=\columnwidth]{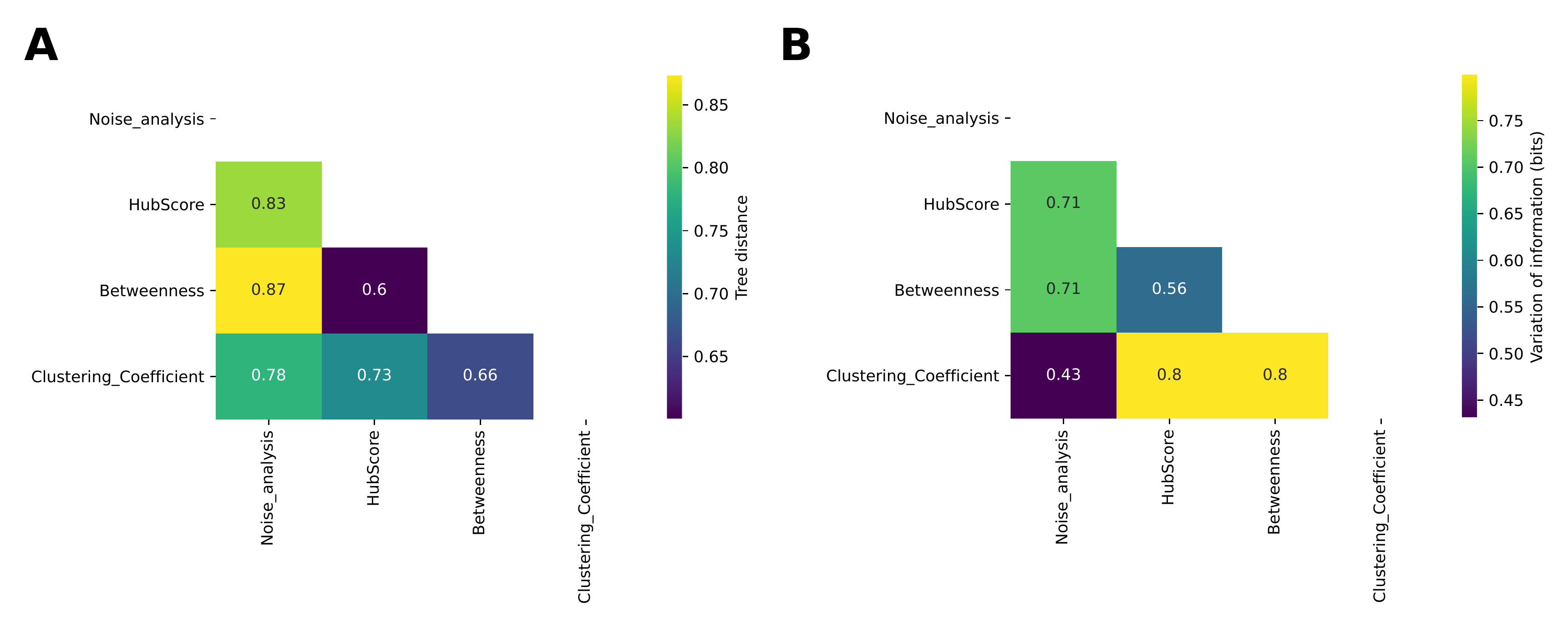}
    \caption{{\bf Comparison between the clusterings.} A: Comparison between the dendrograms of the different hierarchical clustering shown in Figs \ref{fig:clus_net_prop}-\ref{fig:noise} using the Information-based generalized Robinson–Foulds distance (or tree distance). B: Comparison between the partitions obtained using the Variation of Information.}
    \label{fig:clus_comp}
\end{figure}

Finally, we compared the dendrograms obtained from the hierarchical clustering for the different network properties and for the analysis of the noise by computing the Information-based generalized Robinson–Foulds distance, or tree distance (see the Methods section). The results are shown in Fig \ref{fig:clus_comp}A. The most similar dendrograms are those obtained for the Hub score and the betweenness, while the farthest are those obtained for the betweenness and the noise analysis, in agreement with the comparison of the matrices shown in Fig \ref{fig:cmp_matrix}.
The result given by the betweenness and hub score comparison is not surprising, because both descriptors measure the degree of centrality of a node in the complexity of the network. Nonetheless, they are two different types of centrality. The centrality of the betweennwss is expressed in terms of shortest paths, while the hub score centrality is referred to the eigenvalues spectrum of the adjacency matrix. More interesting is the result concerning the difference between the clustering obtained with the betweenness and that obtained with the noise analysis. These two clustering analysis provides the most distant results and this shows how these two corresponding descriptors capture different properties of the system, which influence the final results.
We also compared the partitions obtained by cutting the dendrograms according to the silhouette analysis using the Variation of Information (see the Methods section for details). The most similar partitions are those obtained from the noise analysis and for the clustering coefficient. Indeed, we already noticed that those matrices led to a better separation between oscillating networks and the others. 

%%%%%%%%%%%%%%%%%%%%%%%%%%%%%%%%%%%%%%%%%%
\section{Discussion}

In this manuscript we applied a DA-RNN  to predict the behaviour of stochastic processes. The aim of the work is not only to recover the time traces of the system but also  to infer information on the structure of the system.   A central point of this study is that we used  classical models of gene regulatory network to  define a stochastic system for the simulation of our case studies. Using the in silico data generation approach we  altered  internal parameters of the system whose behaviour was then reproduced through the DNN. This approach gives us full control on the interpretation of the results, preparing the ground for future applications to the output of real experiments. Indeed, state-of-the-art techniques for the measurement of gene expression, such as RNA-seq, can only provide time traces of gene expression, hiding the relationships among the components of the system. The inference of interactions among genes is a critical goal for biology, since understanding the effect of a gene on another allows to predict the behaviour of a gene regulatory network upon perturbations, opening the possibility to design new therapies to reshape the network. 
\\ \\ 
Our work shows that the DA-RNN accurately reconstructs the time traces of genes belonging to different types of gene regulatory networks. In particular, we generated synthetic data from networks with different degree of connectivity (from fully connected to sparser networks), networks displaying oscillatory behaviour or driven by an external signal and networks with a master regulator gene. However, looking at the internal parameters of the attention layer of the neural network, we could not fully reconstruct the internal connection among the genes. Using tools from graph theory, we went beyond this lack of interpretability of the neural network parameters and we showed that considering network properties of the input attention matrices, such as the clustering coefficient, the betweenness centrality and the hub score, it is possible to obtain information about the type of gene regulatory network under study. In particular, the clustering analysis showed that these network properties allow to distinguish different GRN architectures, with the clustering coefficient reflecting better this structure than other properties. We also studied the change in accuracy of the prediction by the neural network under noise addition on protein concentration. Performing a similar analysis, we showed that  the response to noise of GRNs allows to separate different GRN architectures. Moreover, this analysis suggests that the core oscillating genes in oscillatory GRNs are more robust to noise addition, from the point of view of the ability of the neural network in predicting their time traces, compared to the others, while for a network controlled by a master regulator all the genes respond in a similar way.
\\ \\ 
An interesting application of our method is the analysis of time series produced by  RNA-seq data \cite{Selewa2020,Marbach2012}. Given the time series of a gene from a RNA-seq dataset, our work shows that it should be possible to understand to which type of gene regulatory network it belongs, from the properties of the input attention of the DA-RNN used for its prediction. This could be extremely relevant to retrieve information on the biological process in which the gene is involved. Moreover, the large-scale application of our method to all the genes in a dataset could provide a new method, able to predict the future gene dynamic but also to infer regulatory modules.
This work is also different from previous works that used Hopfield neural networks with a symmetric connectivity matrix to model the GRN dynamics storing the observed RNA-seq data as stationary states \cite{Hannam2017,Szedlak2017}, because we actually construct the dynamical model in order to predict the actual dynamics of the system and we use this model to classify different network structures.
\\ \\

In conclusion, we here propose an analysis of gene regulatory networks based on deep neural networks, which exploits the recently introduced DA mechanism and the power of RNN for time series prediction. We were able to reconstruct the  behaviour of the system with high accuracy for most of the GRN architectures. Moreover, through a network analysis of the internal parameters of the input attention we could discriminate the different classes of gene regulatory networks, overcoming the lack of a direct connection between the internal parameters of the DNN and the physical quantities that describe gene interactions. Summing up, our work paves the way to the development of a method for simultaneous time series prediction and analysis of gene interactions from real gene expression data.

%In conclusion, we propose a deep neural network analysis centred on the ideas of attention and recurred neural network that analysing time traces data is able to reconstruct the behaviour of the system with high precision. Even dough the ability of inferring the future of the data point is very accurate for almost all the network taken under consideration, we are not able to directly connect the internal parameter of the DNN with physical quantities, so our method need to be integrated with further analysis. However a matrix analysis on the internal parameters of the DNN is able to discriminate among different Gene Regulatory network class giving a total to understand from the analysis on real data at which kind of network a gene under study belong.

%%%%%%%%%%%%%%%%%%%%%%%%%%%%%%%%%%%%%%%%%%
\vspace{6pt} 

%%%%%%%%%%%%%%%%%%%%%%%%%%%%%%%%%%%%%%%%%%
%% optional
%\supplementary{The following are available online at \linksupplementary{s1}, Figure S1: title, Table S1: title, Video S1: title.}

% Only for the journal Methods and Protocols:
% If you wish to submit a video article, please do so with any other supplementary material.
% \supplementary{The following are available at \linksupplementary{s1}, Figure S1: title, Table S1: title, Video S1: title. A supporting video article is available at doi: link.} 

%%%%%%%%%%%%%%%%%%%%%%%%%%%%%%%%%%%%%%%%%%
\textbf{Author contributions}

Conceptualization, M.M.; methodology, M.M, J.F., E.M. and G.G; software, M.M, J.F., E.M., G.G.; validation, M.M, J.F., E.M. and G.G.; formal analysis, M.M and J.F.; investigation, M.M and J.F.; resources, M.M and J.F.; data curation, M.M and J.F.; writing---original draft preparation, M.M, J.F., E.M., G.G. and G.G.T.; writing---review and editing, M.M, J.F., E.M., G.G. and G.G.T.; visualization, M.M and J.F; supervision, G.G.T.; project administration, M.M. and G.G.T..; funding acquisition, G.G.T.

\medskip

\textbf{Funding}

This research was funded by European Research Council grant  ASTRA number 855923, grant INFORE number 25080. MM was funded by INFORE number 25080.

\medskip

\textbf{Acknowledgments}

The authors thank Antonis Deligiannakins, all members of INFORE, Alexandros Armaos and Alessio Colantoni for interesting discussions.

%\conflictsofinterest{The authors declare no conflict of interest.} 

%% Optional
%\sampleavailability{Samples of the compounds ... are available from the authors.}

%%%%%%%%%%%%%%%%%%%%%%%%%%%%%%%%%%%%%%%%%%
%% Only for journal Encyclopedia
%\entrylink{The Link to this entry published on the encyclopedia platform.}

%%%%%%%%%%%%%%%%%%%%%%%%%%%%%%%%%%%%%%%%%%
%% Optional
%\abbreviations{Abbreviations}{
%The following abbreviations are used in this manuscript:\\
%
%\noindent 
%\begin{tabular}{@{}ll}
%MDPI & Multidisciplinary Digital Publishing Institute\\
%DOAJ & Directory of open access journals\\
%TLA & Three letter acronym\\
%LD & Linear dichroism
%\end{tabular}}

%%%%%%%%%%%%%%%%%%%%%%%%%%%%%%%%%%%%%%%%%%
%% Optional
%\appendixtitles{no} % Leave argument "no" if all appendix headings stay EMPTY (then no dot is printed after "Appendix A"). If the appendix sections contain a heading then change the argument to "yes".
%\appendixstart
\appendix
\section{}
\label{appmeanfield}
%\subsection{}

In this Appendix, we provide the derivation of the mean field equations of the system. We start from the time dependent equation of the probability of having a certain number of protein $n$ of a given species $i$.
 \begin{equation}\label{eqProb}
\dot{P}(n)  =  f_{n-1} P(n-1) + g_{n+1} P(n+1) - P(n) \left[ f_n + g_n \right]~~,
\end{equation}
where $f_n$ and $g_n$ are respectively the rate of probability of increasing or decreasing the number of $n$ of 1. In our system, $f_n$ can be approximated as $f_n \simeq P(x_i^+)dt$, where $P(x_i^+)$ is defined in Eq. \eqref{prob}. We can consider that the proteins decay exponentially such that the $g_n= \alpha n$.

Computing the evolution of the mean value gives
\begin{equation}\label{eqMean}
\dot{ \langle n  \rangle}  = \dot{\left( \sum_n n P(n) \right)} = \sum_n n \dot{P}(n) +  \sum_n P(n) \dot{n} = \sum_n n \dot{P}(n)~~,
\end{equation}
where $\dot{n}$ is 0 because it represents the total number of proteins.
Substituting the terms in the above equation we get
\begin{equation}
\dot{ \langle n  \rangle} =  \sum_n  f_{n-1} P(n-1) + g_{n+1} P(n+1) - P(n) \left[ f_n + g_n \right] n~~,
\end{equation}
where computing the parts containing $f_n$ term by term we get:
\begin{equation}
 f_0 P_0 - f_1 P_1 + 2 f_1 P_1 - 2 f_2 P_2 + 3 f_2 P_2 - 3 f_4 P_4 \dots = \langle f_n \rangle~~.
\end{equation}
For the $g_n$ contributions, computing term by term we find
\begin{equation}
 \alpha \left( \sum_n n (n +1) P(n+1) - n^2 P(n) \right) =  - \alpha \langle n \rangle~~.
\end{equation}
Combining and considering the mean field approximation where $f_{\langle n \rangle} = \langle f_n \rangle$, we can write the equation for the time evolution of the mean value:
\begin{equation}
\dot{\langle n \rangle} = f_{\langle n \rangle} - \alpha \langle n \rangle~~.
\end{equation}
 
Moreover, since we are looking to the concentration dynamics of particles in a growing cell, we need to take into account the dilution effect. To this end, using a classical approach, we assume that, on average, the cell grows following an exponential growth law:
 \begin{equation}
 \frac{d V}{dt } = \lambda V~~.
 \end{equation}
Looking at the dynamic of the concentration of a protein $ \langle x \rangle  =\langle n \rangle /V$  (where $\langle n \rangle $ is the mean number of proteins), the exponential growth assumption
ensures that the dilution term is exactly the growth rate:
 \begin{equation}
 \frac{d \langle x \rangle}{dt} = \frac{d \langle n \rangle /V}{dt}= \frac{d \langle n \rangle }{V dt} - \frac{\langle n \rangle }{V^2}\frac{dV }{dt} =  \frac{\dot{\langle n \rangle }}{V} - \lambda \langle x \rangle~~.
 \end{equation}
Substituting the mean field equation for $n$, we obtain the time evolution dynamic of the average concentration of a protein $x$ in an exponentially growing cell:
 \begin{equation}
\dot{\langle x \rangle} = f_{\langle x \rangle} - (\alpha + \lambda) \langle x \rangle~~.
\end{equation}\label{meanfield}
We note that the $f_{\langle x \rangle}$ term depends on the state of the protein network as described in Eq. \eqref{prob}. The term containing $\alpha $ could also be dependent on the other proteins involved in the system that actively degrade $x$.

\bibliography{biblioGRN}

\end{document}